\documentclass{ws-ijmpd}

\begin{document}

\markboth{Lixin Xu, Hongya Liu} {Three Components Evolution in a
Simple Big Bounce Cosmological Model}

%
\catchline{}{}{}{}{}
%

\title{Three Components Evolution in a Simple Big Bounce Cosmological Model}
\author{Lixin Xu\footnote{lxxu@student.dlut.edu.cn},
Hongya Liu\footnote{hyliu@dlut.edu.cn}}
\address{Department of Physics, Dalian University of Technology,
Dalian, 116024, P. R. China}

\maketitle

\begin{history}
\received{Day Month Year} \revised{Day Month Year}
\end{history}

\begin{abstract}
We consider a five-dimensional Ricci flat Bouncing cosmology and
assume that the four-dimensional universe is permeated smoothly by
three minimally coupled matter components: CDM+baryons $\rho_{m}$,
radiation $\rho_{r}$ and dark energy $\rho_{x}$. Evolutions of
these three components are studied and it is found that dark
energy dominates before the bounce, and pulls the universe
contracting. In this process, dark energy decreases while
radiation and the matter increase. After the bounce, the radiation
and matter dominates alternatively and then decrease with the
expansion of the universe. At present, the dark energy dominates
again and pushes the universe accelerating. In this model, we also
obtain that the equation of state (EOS) of dark energy at present
time is $w_{x0}\approx -1.05$ and the redshift of the transition
from decelerated expansion to accelerated expansion is
$z_{T}\approx 0.37$, which are compatible with the current
observations.
\end{abstract}


\section{Introduction}

In recent decades, the observations of high redshift Type Ia
supernovae reveal that the expansion of our universe is speeding
up rather than slowing down \cite{RS}\cite{TKB}\cite{Riess}.
Meanwhile, the discovery of Cosmic Microwave Background (CMB)
anisotropy on degree scales indicates $\Omega _{total}\simeq 1$
\cite{BHS}, and the galaxy redshift surveys indicates $\Omega
_{m}\simeq \left. 1\right/ 3$. All these strongly suggest that the
universe is permeanted smoothly by 'dark energy', which violate
the strong energy condition and has negative pressure. The dark
energy and accelerating universe has been discussed extensively
from different points of view. Usually, inspired by inflation,
dark energy was treated as a scalar field which is minimally
coupled with conventional matter, such as in quintessence
\cite{Quintessence}, phantom \cite{Phantom} and k-essence
\cite{K-essence} model. The kinematic interpretation of the
relationship between SN Ia luminosity distance and red-shift
implies that the transition from decelerated expansion to
accelerated expansion is around $z_{T}=0.46\pm 0.13$ \cite{Riess}.

It has been drawn great attention to the idea that our conventional universe
is embedded in a higher dimensional world as required in the Kaluza-Klein
theories and the brane world theories. In this paper, we consider
a class of five-dimensional cosmological model which, as an alternative candidate to
the standard $4D$ FRW model, has been discussed by many authors \cite%
{LWX}. Instead of the Big Bang singularity of the standard model,
this $5D$ cosmological model is characterized by a 'Big Bounce',
which corresponds to a finite and minimal size of the universe.
Before the bounce the universe contracts, and after the bounce it
expands. This model is $5D$ Ricci-flat, as in Space-Time-Matter
(STM) theory \cite{Wesson}, implying that the $5D$ space-time is
empty, and the matter of the conventional $4D$ universe is induced
from the fifth dimension. This approach is guaranteed by
Campbell's theorem \cite{Compbell} that any solution of Einstein
equation of $N$ dimensions can be locally embedded in Ricci-flat
manifold of $\left( N+1\right) $ dimensions. So, the theory is
consistent with the general theory of relativity (GR) locally.
But, at large scale, it maybe different from GR. This is the
motivation of this paper. In a previous work \cite{LWX}, a time
variable cosmological 'constant ' is isolated out in a natural way
from the induced $4D$ energy-momentum tensor. In this paper,
instead of isolating a cosmological 'constant' from the energy
momentum tensor, we assume that the universe is permeated smoothly
by three components: CDM+baryons $\rho _{m}$ with pressure
$p_{m}=0$, radiation $\rho_{r}$ with pressure
$p_{r}=\rho_{r}\left/ 3\right.$, and dark energy $\rho_{x}$ with
pressure $p_{x}=w_{x}\rho_{x}$ ($w_{x}$ is in general a function
of time, which is not put by prior.). By studying we will find
that dark energy dominates before the bounce and pulls the
universe contracting. In this process, dark energy decreases and
the radiation and the matter increase. After the bounce, the
radiation and the matter dominates alternatively and then decrease
with the expansion of the universe. At present stage, dark energy
dominates again and pushes the universe accelerating.

\section{The dimensionless density parameters of the three components in the
$5D$ model}

Within the framework of STM theory, an exact $5D$ cosmological solution was
given by Liu and Mashhoon in 1995 \cite{Liu}. Then, in 2001, Liu and Wesson
\cite{LWX} restudied the solution and showed that it describes a
cosmological model with a big bounce as opposed to a big bang. The $5D$
metric of this solution reads
\begin{equation}
dS^{2}=B^{2}dt^{2}-A^{2}\left( \frac{dr^{2}}{1-kr^{2}}+r^{2}d\Omega
^{2}\right) -dy^{2},  \label{5-metric}
\end{equation}%
where $d\Omega ^{2}\equiv \left( d\theta ^{2}+\sin ^{2}\theta d\phi
^{2}\right) $ and
\begin{eqnarray}
A^{2} &=&\left( \mu ^{2}+k\right) y^{2}+2\nu y+\frac{\nu ^{2}+K}{\mu ^{2}+k},
\nonumber \\
B &=&\frac{1}{\mu }\frac{\partial A}{\partial t}\equiv \frac{\dot{A}}{\mu }.
\label{A-B}
\end{eqnarray}%
Here $\mu =\mu \left(t\right)$ and $\nu =\nu (t)$ are two arbitrary functions of $t$, $%
k $ is the $3D$ curvature index $\left( k=\pm 1,0\right) $, and $K$ is a
constant. This solution satisfies the 5D vacuum equations $R_{AB}=0$. So we
have three invariants
\begin{equation}
I_{1}\equiv R=0,I_{2}\equiv R^{AB}R_{AB}=0,I_{3}=R_{ABCD}R^{ABCD}=\frac{%
72K^{2}}{A^{8}},  \label{3-invar}
\end{equation}%
which show that $K$ determines the curvature of the $5D$ manifold. Using the
$4D$ part of the $5D$ metric (\ref{5-metric}) to calculate the $4D$ Einstein
tensor, one obtains
\begin{eqnarray}
^{(4)}G_{0}^{0} &=&\frac{3\left( \mu ^{2}+k\right) }{A^{2}},  \nonumber \\
^{(4)}G_{1}^{1} &=&^{(4)}G_{2}^{2}=^{(4)}G_{3}^{3}=\frac{2\mu \dot{\mu}}{A
\dot{A}}+\frac{\mu ^{2}+k}{A^{2}}.  \label{einstein}
\end{eqnarray}

It can be seen \cite{LWX} that there are two kinds of
singularities corresponding to $A=0$ and $B=0$, respectively.
$A=0$ represents the usual "Big Bang" singularity. $B=0$ (with $A
\ne 0$) represents a new kind of singularity at which the three
invariants in (\ref{3-invar}) are regular while $A$ reaches to its
minimum ($B=\dot{A}\left/\mu\right.$). So this new kind of
singularity corresponds to a bouncing.

In the previous work \cite{LWX}, the induced matter was assumed
that to be a conventional matter plus a cosmological 'constant'
(which in fact is not constant but a function of time). In this
paper, we assume the induced matter contains CDM+baryons $\rho
_{m}$, radiation $\rho _{r}$ and dark energy $\rho _{x}$, which
are minimally coupled with each other. So, we have
\begin{eqnarray}
\frac{3\left( \mu ^{2}+k\right) }{A^{2}} &=&\rho _{m}+\rho _{r}+\rho _{x},
\nonumber \\
\frac{2\mu \dot{\mu}}{A\dot{A}}+\frac{\mu ^{2}+k}{A^{2}}
&=&-p_{m}-p_{r}-p_{x},  \label{FRW-Eq}
\end{eqnarray}%
with
\begin{eqnarray}
p_{m} &=&0,p_{r}=\rho _{r}/3,  \label{EOS-M} \\
p_{x} &=&w_{x}\rho _{x}.  \label{EOS-X}
\end{eqnarray}

From Eqs. (\ref{FRW-Eq}), (\ref{EOS-M}) and (\ref{EOS-X}), we obtain the
equation of state (EOS) of the dark energy
\begin{equation}
w_{x}=\frac{p_{x}}{\rho _{x}}=-\frac{2\left. \mu \dot{\mu}\right/ A\dot{%
A}+\left. \left( \mu ^{2}+k\right) \right/ A^{2}+\left. \rho
_{r0}A^{-4}\right/ 3}{3\left. \left( \mu ^{2}+k\right) \right/ A^{2}-\rho
_{m0}A^{-3}-\rho _{r0}A^{-4}}  \label{wx}
\end{equation}%
and the dimensionless density parameters
\begin{eqnarray}
\Omega _{m} &=&\frac{\rho _{m}}{\rho _{m}+\rho _{r}+\rho _{x}}=\frac{\rho
_{m0}}{3\left( \mu ^{2}+k\right) A},  \label{omiga-M} \\
\Omega _{r} &=&\frac{\rho _{r}}{\rho _{m}+\rho _{r}+\rho _{x}}=\frac{\rho
_{r0}}{3\left( \mu ^{2}+k\right) A^{2}},  \label{omiga-R} \\
\Omega _{x} &=&1-\Omega _{m}-\Omega _{r}.  \label{omiga-X}
\end{eqnarray}%
where $\rho _{m0}=\bar{\rho}_{m0}A_{0}^{3}$, $\rho _{r0}=\bar{\rho}%
_{r0}A_{0}^{4}$ denote the present volume of the matter and radiation, $\bar{%
\rho}_{m0}$, $\bar{\rho}_{r0}$ is the current value of CDM+baryons
and radiation densities, respectively.

\section{The evolution of the three components in the $5D$ model}

The Eqs. (\ref{wx})-(\ref{omiga-X}) contain only two arbitrary
functions $\mu \left( t\right) $, $\nu \left( t\right) $ and
another two parameters $K$ and $y$. So, by choosing the function
$\mu \left( t\right) $, $\nu \left( t\right) $ and the parameters
$K$ and $y$ properly, we can obtain properties of the dark energy
which coincide with the present astronomical observing data. Under
these choices, we discuss the evolution of the three components.
Because the observations support a flat universe, we only consider
the case $k=0$. In this special case, the EOS of the dark energy
and the dimensionless density parameters become
\begin{eqnarray}
w_{x} &=&-\frac{2\left. \mu \dot{\mu}\right/ A\dot{A}+\left. \mu
^{2}\right/ A^{2}+\left. \rho _{r0}A^{-4}\right/ 3}{3\left. \mu
^{2}\right/
A^{2}-\rho _{m0}A^{-3}-\rho _{r0}A^{-4}},  \label{new-omiga-x} \\
\Omega _{m} &=&\frac{\rho _{m0}}{3\mu ^{2}A},\Omega _{r}=\frac{\rho _{r0}}{%
3\mu ^{2}A^{2}},\Omega _{x}=1-\Omega _{m}-\Omega _{r}.  \label{new-omiga}
\end{eqnarray}%
For the complexity of the solutions, we analyze the properties of
the solutions numerically.

Now, we analyze the dimensionless density parameters and compare
them with
the observed data. From Eq. (\ref{omiga-M})-Eq. (\ref{omiga-X}), using $%
A_{0}\left/ A\right. =1+z$ and $\Omega _{m}\left/ \Omega
_{x}\right. =\gamma_{z}$, we obtain the relation
\begin{equation}
\frac{3\mu_{z} ^{2}}{A_{0}^{2}}=\left(
1+\frac{1}{\gamma_{z}}\right) \bar{\rho}_{m0}\left( 1+z\right)
+\bar{\rho}_{r0}\left( 1+z\right) ^{2}.  \label{u}
\end{equation}%
So, at present time $z=0$, and at the transition time from
decelerated to accelerated expansion we let $z=z_{T}$. Then
(\ref{u}) the relation gives
\begin{eqnarray}
\frac{3\mu _{0}^{2}}{A_{0}^{2}} &=&\left( 1+\frac{1}{\gamma
_{0}}\right) \bar{\rho
}_{m0}+\bar{\rho}_{r0},  \label{present} \\
\frac{3\mu _{T}^{2}}{A_{0}^{2}} &=&\left( 1+\frac{1}{\gamma
_{T}}\right) \bar{\rho}_{m0}\left(
1+z_{T}\right)+\bar{\rho}_{r0}\left( 1+z_{T}\right) ^{2}.
\label{transition}
\end{eqnarray}%
At the equilibrium point $z_{E}$, the matter density equals to the
radiation density. So we have $\bar{\rho}_{r0}\sim
\bar{\rho}_{m0}\left/\left(1+z_{E}\right) \right.$, where
$z_{E}\sim 6000$. Thus we obtain
\begin{equation}
\frac{\mu _{0}^{2}}{\mu _{T}^{2}}=\frac{1+1\left/ \gamma
_{0}\right. }{\left( 1+1\left/ \gamma _{T}\right. \right) \left(
1+z_{T}\right)}.  \label{def-condition}
\end{equation}
The observed value $\gamma_{0}$ is $\gamma_{0}\sim
3\left/7\right.$ at present $z=0$, and $\gamma_{T}\sim 1$ at the
transition $z=z_{T}$. So, we obtain
\begin{equation}
\frac{\mu_{0}^2}{\mu_{T}^{2}}\sim \frac{5}{3\left(1+z_{T}\right)}.
\label{0-T}
\end{equation}
This implies that the function $\mu\left(t \right)$ determines the
transition from decelerated expansion to accelerated expansion at
the late epoch of the universe. In addition, the other function
$\nu\left(t \right)$ would be constrained by $z_{E}$.

The Hubble and deceleration parameters should be given as
\cite{Liu-b}
\begin{eqnarray}
H \left(t, y \right) \equiv \frac{1}{A} \frac{dA}{d\tau}
=\frac{1}{B}\frac{\dot{A}}{A}=\frac{\mu}{A}, \nonumber \\
q \left(t, y\right)\equiv \left.
-A\frac{d^{2}A}{d\tau^{2}}\right/\left(\frac{dA}{d\tau}\right)^{2}
=-\frac{A \dot{\mu}}{\mu \dot{A}}, \label{decf}
\end{eqnarray}
from which we see that $\dot{\mu}\left/\mu\right.>0$ represents an
accelerating universe, $\dot{\mu}\left/\mu\right.<0$ represents a
decelerating universe. So the function $\mu(t)$ plays a crucial
role of defining the properties of the universe in late time
again. The deceleration parameter $q_{T}=0$ $(\dot{\mu}=0)$
corresponds to the transition from deceleration to acceleration.

The scale factor $A$ in Eq. (\ref{A-B}) can also be written in the
form
\begin{equation}
A^{2}=\frac{1}{\mu^{2}+k}\left[\left(\mu^{2}+k\right)y+\nu\right]^{2}+\frac{K}{\mu^{2}+k}.
\label{new-A}
\end{equation}
The three values of $K$ ($K>0, =0, <0$) represents three types of
the $5D$ manifold. By rescaling $\mu^{2}$ and $k$, we can set
these types as $K=+1, 0, -1$. In this paper we consider the type
$K=1$ which corresponds to a bouncing universe. Meanwhile, from
(\ref{new-A}) we can also see that the form $A(t,y)$ is invariant
under a translation along the $y$-direction $y \rightarrow
y+y_{0}$ provided we redefine $\nu(t) \rightarrow
\nu(t)-(\mu^{2}+k)y_{0}$. Therefore, we can set $y=1$ without lose
of generality. So, in Eq. (\ref{new-A}), we only have to
determinate $\mu(t)$ and $\nu(t)$. Supposing
$\mu\left(t\right)=at+b\left/t\right.$, $\nu(t)=ct$. The three
constants $a, b$ and $c$ should be constrained and determined by
the observations, such as the dimensionless density parameters
$\Omega_{m0}$, the EOS of dark energy $w_{x0}$, the decelerated
factor $q_{0}$ and the transition redshift $z_{T}$.

\begin{figure}[th]
\centerline{\psfig{file=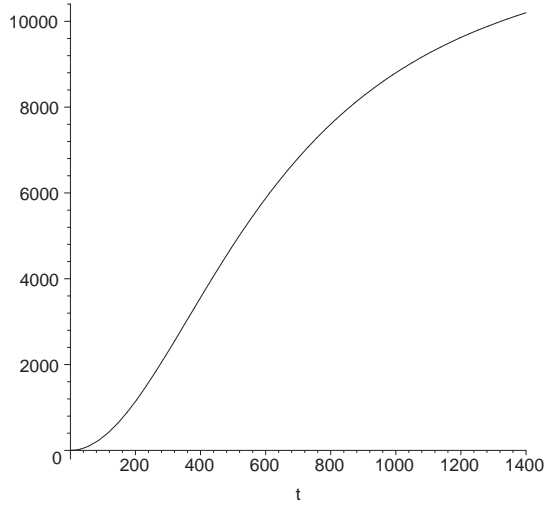,width=7cm}} \vspace*{8pt}
\caption{The evolution of the scale factor $A\left( t,y\right) =\protect%
\sqrt{\left( \protect\mu y+\protect\nu \left/ \protect\mu \right.
\right) ^{2}+K\left/ \protect\mu ^{2}\right. }$ with time $t\in(0,
14)$. Here, $K=1$, $y=1$, $\rho _{m0}=1.1$, $\rho _{r0}=2.4$,
$a=0.000009,$ $b=3.5$, $ c=0.11$. The bounce can be seen from the
figure.} \label{f1-1}
\end{figure}

\begin{figure}[th]
\centerline{\psfig{file=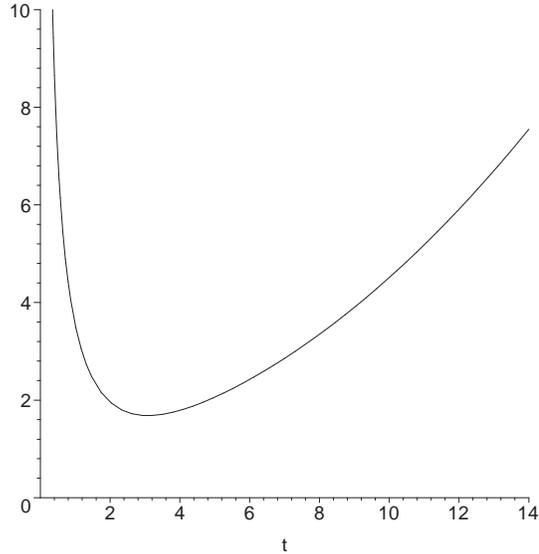,width=7cm}} \vspace*{8pt}
\caption{The evolution of the scale factor $A\left( t,y\right) =\protect%
\sqrt{\left( \protect\mu y+\protect\nu \left/ \protect\mu \right.
\right) ^{2}+K\left/ \protect\mu ^{2}\right. }$ with time $t\in(0,
1000)$. Here, $K=1$, $y=1$, $\rho _{m0}=1.1$, $\rho _{r0}=2.4$,
$a=0.000009,$ $b=3.5$, $ c=0.11$.} \label{f1-2}
\end{figure}

\begin{figure}[th]
\centerline{\psfig{file=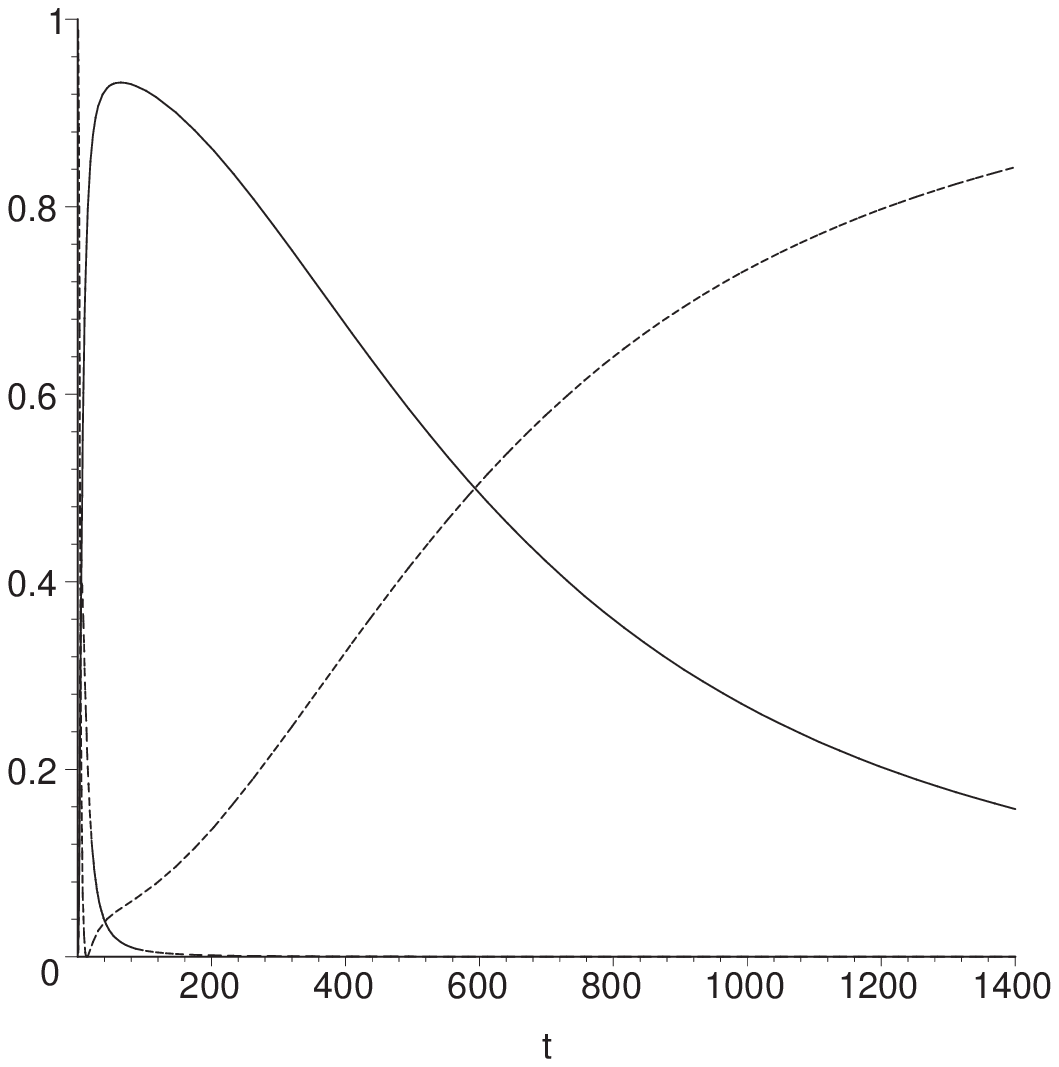,width=7cm}} \vspace*{8pt}
\caption{The evolution of the three components $\Omega _{m}$ (the
solid line), $\Omega _{r}$ (the dashed line), $\Omega _{x}$ (the
dotted line). Where, $K=1$, $y=1$, $\rho _{m0}=1.1$, $\rho
_{r0}=2.4$, $a=0.000009,$ $b=3.5$, $ c=0.11$. The redshift of
transition from decelerated expansion to accelerated expansion is
$z_{T}\sim 0.37$.} \label{f2}
\end{figure}

\begin{figure}[th]
\centerline{\psfig{file=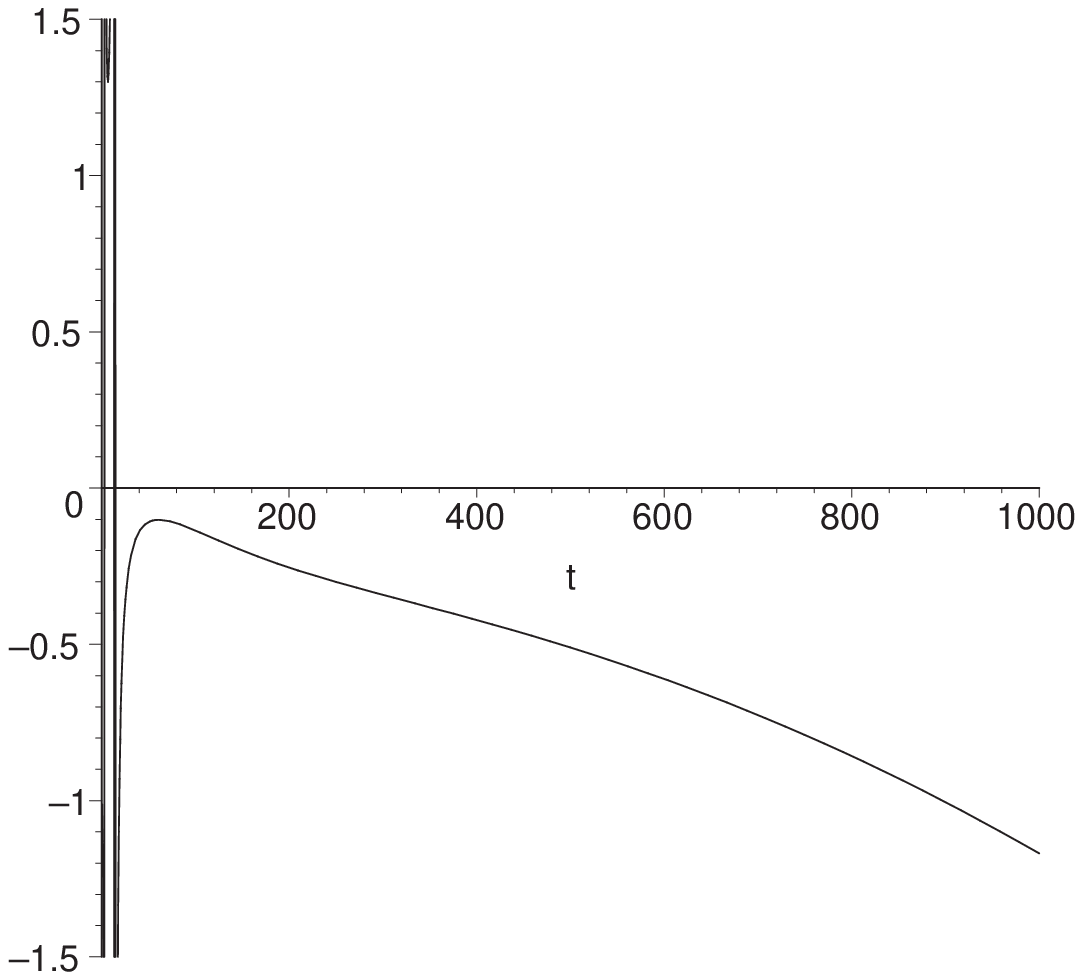,width=7cm}} \vspace*{8pt}
\caption{The EOS of dark energy evolution with time $t$. Here,
$K=1$, $y=1$, $\rho _{m0}=1.1$, $\rho _{r0}=2.4$, $a=0.000009,$
$b=3.5$, $ c=0.11$.} \label{f3}
\end{figure}

\begin{figure}[th]
\centerline{\psfig{file=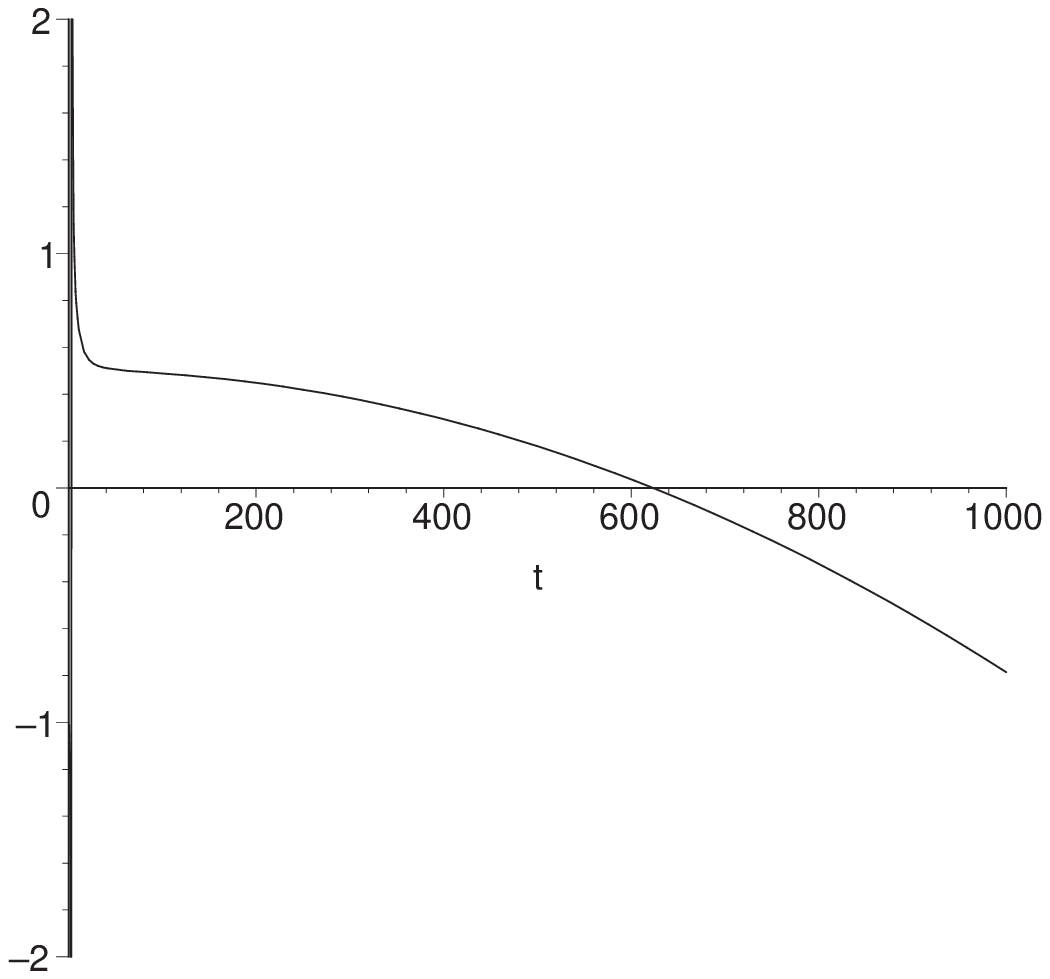,width=7cm}} \vspace*{8pt}
\caption{The evolution of deceleration parameter with time t.
Here, $K=1$, $y=1$, $\rho _{m0}=1.1$, $\rho _{r0}=2.4$,
$a=0.000009,$ $b=3.5$, $ c=0.11$. The time $t$ and redshift of
transition from decelerated expansion to accelerated expansion are
$t_{T}=\sqrt{b\left/a\right.}\sim 624$ and $z_{T}\sim 0.37$
respectively.} \label{f4}
\end{figure}

By the above choice, we can obtain the transition time
$t_{T}=\sqrt{b\left/a\right.}$ from Eq. (\ref{decf}). It is to say
that the ratio of the parameters $b$ and $a$ constrain the
transition from deceleration to acceleration. The model
independent estimation of the cosmological parameters values at
the present is $\Omega_{m0}\sim 0.3$, $\Omega_{x0}\sim 0.7$. To
meet these data, we choose $K=1$, $y=1$, $\rho _{m0}=1.1$, $\rho
_{r0}=2.4$, $a=0.000009,$ $b=3.5$, $ c=0.11$, the evolution of the
scale factor $A$ is plotted in Fig. \ref{f1-1} and Fig.
\ref{f1-2}.
The evolution of the three components is plotted in Fig. \ref{f2}.
The present observed values are $\Omega _{m}\approx 0.3$, $\Omega
_{x}\approx 0.7$. We can see that they correspond to the time
$t_{0}\approx 928$ in Fig. \ref{f2}. Meanwhile, $\Omega
_{m}=\Omega _{x}$ corresponds to the time $t\approx 595$. So from
Fig. \ref{f1-2}, we can conclude that the redshift of the
transition from deceleration to acceleration is $z_{T}\approx
0.37$, which is close to the observation in \cite{Riess}. In Fig.
\ref{f1-1}, the bounce time is $t_{b}\approx 3$. So in Fig.
\ref{f2}, we can see that: Before the bounce, the dark energy
dominates and pull the universe to contract. After the bounce, the
radiation and CDM+baryons dominate alternatively. The radiation
dominates firstly, then the CDM+baryons dominated. After $t
\approx 624$, the dark energy dominates again and push the
universe accelerating at present. Also, the EOS of dark energy is
plotted in Fig. \ref{f3}. From above calculation, we can read
$w_{x0}=-1.05$ at present.
The evolution of the deceleration factor with $t$ is plotted in
Fig. \ref{f4}. The universe begins accelerating from $z_{T}$ on.
From the Fig. \ref{f4}., the transition time from deceleration to
acceleration is $t_{T} \sim 624$, which corresponds to the
redshift is $z_{T} \approx 0.37$

\section{Conclusions}

The classes of five-dimensional cosmological solution
(\ref{5-metric}) is characterized by a 'Big Bounce' which
contrasted with the 'Big Bang' in standard cosmological models.
Mathematically, the solution contains two arbitrary functions $\mu
\left( t\right) $, $\nu \left( t\right)$. Different choices of the
functions may give different models to describe different stages
of the universe evolution. In this paper, the induced matter
contains three components $\rho _{m}$, $\rho _{r}$, $\rho _{x}$.
By choosing the two arbitrary functions properly, we conclude that
before the bounce, the dark energy dominates and pull the universe
to contract. After the bounce, the radiation and CDM+baryons
dominate alternatively. Firstly the radiation dominates, then the
CDM+baryons dominates. At not a distance past, the dark energy
dominates again and push the universe accelerating. The equation
of state of dark energy is $w_{x0}\approx -1.05$ at present. The
redshift of the transition from decelerated expansion to
accelerated expansion is $z_{T}\approx 0.37$. These results are
compatible with the current observations.

\section*{Acknowledgments}This work was supported by NSF (10273004)  and NBRP
(2003CB716300) of P. R. China.

\end{document}